\documentclass[preprint,aps,tightenlines]{revtex4}
\usepackage[dvipdf]{graphicx}

\begin{document}
\title{Atom interferometry}

\author{A. Miffre, M. Jacquey, M. B\"uchner, G. Tr\'enec and J. Vigu\'e}
\address{Laboratoire Collisions, Agr\'egats, R\'eactivit\'e (UMR 5589 CNRS-UPS),
\\ IRSAMC, Universit\'e Paul Sabatier Toulouse 3, 31062 Toulouse
cedex 9, France
\\ e-mail:~{\tt jacques.vigue@irsamc.ups-tlse.fr}}

\date{\today}

\begin{abstract}

In this paper, we present a brief overview of atom interferometry.
This field of research has developed very rapidly since 1991. Atom
and light wave interferometers present some similarities but there
are very important differences in the tools used to manipulate
these two types of waves. Moreover, the sensitivity of atomic
waves and light waves to their environment is very different. Atom
interferometry has already been used for a large variety of
studies: measurements of atomic properties and of inertial effects
(accelerations and rotations), new access to some fundamental
constants, observation of quantum decoherence, etc. We review the
techniques used for a coherent manipulation of atomic waves and
the main applications of atom interferometers.


\end{abstract}

\maketitle


\section{Introduction}

Several atom interferometers
\cite{carnal91,keith91,riehle91,kasevich91} gave their first
signals in 1991 and atom interferometry has developed very rapidly
since. Some review papers have already appeared
\cite{adams94,baudon99} and the book "Atom interferometry" edited
by Berman \cite{berman97} in 1997 represents an excellent
introduction to this field. The present paper gives an overview of
atom interferometry. After a comparison of atomic and light waves,
we describe the sources of atomic waves and the tools used for
their coherent manipulation. We then present the main types of
interferometers and their applications. The non-linear effects due
to atom-atom interactions (recently reviewed by Bongs and
Sengstock \cite{bongs04}) are not discussed here.

\section{Main new features of atom interferometry}

The main differences between light waves and atomic waves are
their dispersion relations and their group velocities. As we
neglect here atom-atom interactions in the atomic wave, we may
consider only single-atom plane waves described by a ket $\left|
{\mathbf k}, i\right>$, where ${\mathbf k}$ is the wave vector and
$i$ is the atom internal state which replaces the polarization
vector ${\mathbf {\varepsilon}}$ of a light wave. The total energy
$E_{tot} = \hbar \omega $ of such a state is the sum of the
internal energy $E_i$ (including the rest mass energy $mc^2$) and
the kinetic energy given, in the non-relativistic limit, by
$E_{kin} = \hbar^2 {\mathbf k}^2/(2m)$:
\begin{equation}
\label{n0} \hbar \omega = E_i + \hbar^2 \frac{{\mathbf k}^2}{2m}
\end{equation}
\noindent from which we get the group velocity equal to the
classical velocity ${\mathbf v}$:

\begin{equation}
\label{n1} \partial\omega /\partial {\mathbf k} = \hbar {\mathbf
k}/m = {\mathbf v}
\end{equation}
\noindent The dependence of the group velocity on $k$ induces the
well-known wave packet spreading: vacuum is dispersive for matter
waves while it is not for light. From a practical point of view, a
very important quantity is the de Broglie wavelength given by:

\begin{equation}
\label{n2} \lambda_{dB}= \frac{2\pi}{k} = \frac{h}{mv} \approx
\frac{4\times 10^{-7}}{A v} \mbox{  meter}
\end{equation}

\noindent where $A$ is the mass number and $v$ is the velocity in
meters per second. For thermal atoms or molecules with $ v \sim
10^3$ m/s, the de Broglie wavelength is $\lambda_{dB} \sim
(0.4/A)$ nanometers. For cold and ultra-cold atoms, with
velocities in the millimeter/second to meter/second range, the de
Broglie wavelength may be comparable to $1$ micrometer or even
larger.

Finally, the sensitivity of atomic waves to inertial effects is
considerably larger than the one of light waves and this is a
consequence of their considerably smaller group velocity: during
the time spent by an atom inside an interferometer, a rotation or
an acceleration changes the lengths of the interfering paths, thus
inducing a phase shift of the interference signals.

\section{Main tools of atom interferometry}

\subsection{Sources of atomic waves and detectors}

The simplest source is a thermal atomic beam, either effusive or
supersonic, this last type providing a narrower velocity
distribution and, in both cases, the flux at the interferometer
output is very small. Very efficient detectors are needed and most
experiments with thermal atoms have been done either with alkali
or with metastable atoms which can be very efficiently detected
using surface ionization.

Cold atoms give access to very long interaction times, which
improves the ultimate resolution of a measurement. Therefore, many
experiments use laser-cooled gases with the following scheme: an
atomic trap is first loaded; after a final cooling step and
optical pumping in a particular sub-level, the gas cloud is
accelerated by laser beams and sent into the interferometer. The
usual detection technique is then based on laser fluorescence,
which allows the selective measurement of the populations of the
ground state hyperfine levels. In most cases, the interference
signal is to be found in the repartition of the population among
these hyperfine levels. The possibility of measuring on each
atomic cloud the populations of two hyperfine levels reduces the
noise, down to the quantum projection limit
\cite{itano93,santarelli99}, well below the fluctuations from shot
to shot.

Degenerate quantum gases, Bose Einstein condensate (BEC) or Fermi
degenerate gases, can be used as sources for atom interferometers
and the atom-laser beams extracted from BEC are analogous to laser
beams in optics. For the high densities achieved in BEC, the
atom-atom interactions are not negligible and, at the present
state of the art, these sources are especially interesting for
non-linear atom optics (as reviewed by \cite{bongs04}). Only some
experiments, in which these non-linear aspects are weak, will be
discussed here.

\subsection{Coherent manipulation of atomic waves}

\subsubsection{Diffraction by material structures}

Diffraction of atoms by crystal surfaces, first observed by Stern
and Estermann in 1929-1930 \cite{stern29,estermann30}, is used
nowadays to study surface order and surface excitations. This
diffraction process has not been used for an atom interferometer,
because of the extreme requirements on surface quality and
positioning.

Diffraction by material slits is obviously possible and a Young's
double slit experiment was realized by Carnal and Mlynek
\cite{carnal91} in 1991.

Diffraction by a grating made of nanowires is also possible and
high quality gratings with periods down to $100$ nm can be made by
nanolithography techniques \cite{ekstrom92,savas95} with areas
close to $1$ mm$^2$. These gratings can diffract any atom,
molecule or cluster \cite{schollkopf04}. Sch\"ollkopf and Toennies
\cite{schollkopf94} have used diffraction of a supersonic beam as
a mass selection process for weakly bound helium clusters.

As discussed below, the atom-surface van der Waals interaction
cannot be neglected. The use of gratings and nanostructures with
cold atoms is not common, probably because of the strength of the
atom-grating van der Waals interaction. However, Shimizu and
co-workers have developed  atom holograms in transmission
\cite{morinaga96a,fujita00} and also in reflection
\cite{shimizu02}, using the quantum reflection regime.

\subsubsection{Diffraction by a laser standing wave}

In 1933, Kapitza and Dirac \cite{kapitza33} proposed to diffract
an electron beam by a standing light wave, in order to prove the
existence of stimulated emission of radiation but this experiment
was feasible only with lasers \cite{batelaan01}. In 1966,
Altshuler et al. \cite{altshuler66} extended this idea to the
diffraction of atoms: the diffraction probability was predicted to
be considerably larger, especially if the laser frequency is close
to a resonance transition.

\begin{figure}
\includegraphics[width = 8 cm,height= 6cm]{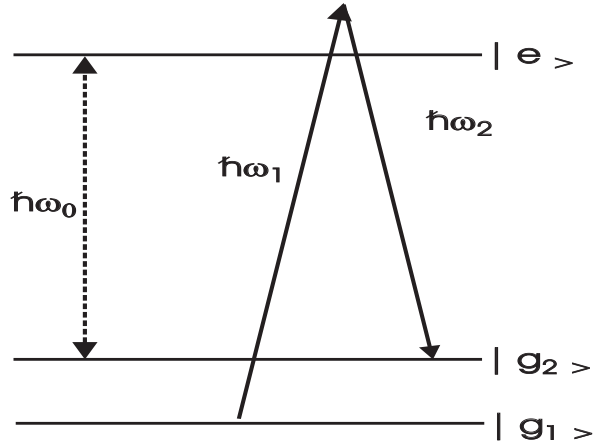}
\caption{\label{f1} Principle of atom diffraction by laser using a
Raman process: the atom absorbs a photon of energy
$\hbar\omega_{1}$ and emits a photon of energy $\hbar\omega_{2}$
by stimulated emission, thus making a transition from state
$\left| g_1 \right>$ to state $\left| g_2 \right>$. The case of
elastic diffraction is deduced from the Raman case by making
$\left| g_1 \right> = \left| g_2 \right>$ and $\hbar\omega_{1} =
\hbar\omega_{2}$. In this last case, diffraction of order $p$ can
be observed, as the absorption-stimulated emission cycle may occur
$p$ times.}
\end{figure}

\begin{figure}
\includegraphics[width = 9 cm,height= 4 cm]{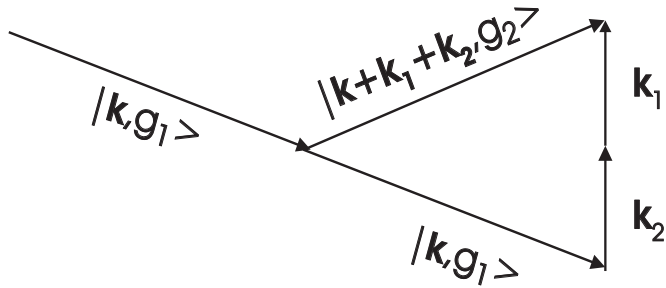}
\caption{\label{f2} Momentum conservation in atom diffraction by
laser: the atom receives a momentum kick equal to $\hbar {\mathbf
k}_1$ due to the absorption of a photon of energy
$\hbar\omega_{1}$ going in one direction and a second kick equal
to $\hbar {\mathbf k}_2$ in the same direction, due to the
stimulated direction of a photon of energy $\hbar\omega_{2}$ going
in the opposite direction. The conservation of the atom kinetic
energy requires the Bragg geometry, as illustrated in this figure.
As in Fig. 1, elastic diffraction is deduced from the Raman case
by making  $\hbar {\mathbf k}_1 = \hbar {\mathbf k}_2$.}
\end{figure}

During a diffraction process (see Fig. 1), the atom absorbs a
photon $\left| \omega, {\mathbf k} \right>$ going in one direction
and makes a stimulated emission of a photon $\left| \omega,
-{\mathbf k}\right>$ going in the opposite direction. If the two
photons have the same polarization, this process is fully elastic,
i. e. the initial and final internal states are identical.
Conservation of energy and momentum is exactly fulfilled in the
Bragg geometry, as shown in Fig. \ref{f1} and \ref{f2}. Usually,
the laser standing wave has a finite spatial width (respectively a
finite duration): the corresponding dispersion of the photon
momentum around its mean value ${\mathbf k}$ (respectively of its
frequency around its mean value $\omega$) relaxes the conservation
laws. In the simplest case, laser diffraction can be described as
a coherent evolution of the atom among two states differing by
their momentum and this process is therefore a Rabi oscillation.

A standing wave moving in the laboratory with a velocity $v$ can
be produced by using two counter-propagating laser beams with
different frequencies (the velocity being proportional to the
frequency difference). This possibility is widely used to
characterize the momentum distribution of ultra-cold atoms
\cite{bongs04,rolston02}.

The calculation of the diffraction probability has been the
subject of many works, corresponding to limiting cases, such as
the Raman-Nath case (thin grating) \cite{bernhardt81} or the Bragg
case (thick grating and weak potential) \cite{giltner95a}. The
various possible regimes are discussed in reference
\cite{keller99}. Bloch states, which have been introduced by
Letokhov and Minogin \cite{letokhov81} (see also \cite{castin91})
to describe the motion of the atom in a standing wave provide an
unified treatment of atom diffraction \cite{champenois01}.

The photon recoil effect was first observed in a saturated
absorption spectroscopy experiment by Hall and Bord\'e
\cite{hall76} in 1976. Resolved diffraction peaks were observed
\cite{moskowitz83} in 1983 and Bragg scattering \cite{martin88} in
1988, both by Pritchard and co-workers. Siu Au Lee and co-workers
\cite{giltner95a} were able to observe Bragg diffraction up to the
sixth order and to build an interferometer operating with any
diffraction order $p$ from $p=1$ to $3$ \cite{giltner95b}.

This diffraction process can be generalized to several cases as
explained by Bord\'e \cite{borde89,borde97}. We will discuss here
only the case of Raman diffraction (see Fig. \ref{f1} and
\ref{f2}). The atom has two ground state hyperfine sub-levels
$\left| g_1 \right>$ and $\left| g_2 \right>$, with an energy
splitting $\hbar \omega_{12}$. The laser standing wave is replaced
by two counter-propagating waves of frequencies $\omega_{1}$ and
$\omega_{2}$, with $\omega_{1}- \omega_{2}= \omega_{12}$. The
diffraction corresponds to the absorption of a photon $\omega_{1}$
and the stimulated emission of a photon $\omega_{2}$, while the
atom makes a Raman transition from state $\left|g_1\right>$ to
state $\left|g_2 \right>$. The main advantage of such an inelastic
diffraction process is that the transmitted and diffracted beams
differ by their internal states. It is therefore very easy to
detect selectively the direct and diffracted beams, but this
diffraction process is coherent only if the laser beat note
$(\omega_{1}- \omega_{2})$ is phase-locked on a stable oscillator.

Finally, we have not discussed the limitations of this diffraction
process. The laser frequencies are usually chosen close to
resonance but not exactly at resonance, so that the probability of
a spontaneous photon emission remains negligible. A different
regime is based on adiabatic transfer and, then, the laser is
exactly at resonance \cite{marte91}. In this case, the transfer of
a very large number of photon momenta has been demonstrated
\cite{weitz94}.

Diffraction by laser has a very important advantage with respect
to diffraction by material gratings: the diffraction amplitude can
be rapidly modulated as a function of time and this possibility is
widely used to build temporal interferometers.

\subsubsection{Mirrors, traps and microtraps}

A repulsive potential can be used to build mirrors for atomic
waves:

- the Earth gravitational potential reflects an atomic beam going
upward, producing an atomic fountain.

- laser evanescent waves with a positive detuning ($\omega >
\omega_0$) have been used as mirrors \cite{aminoff93}. The atom
feels the sum of the van der Waals attractive potential of the
surface and the dipole repulsive potential due to the evanescent
wave \cite{landragin96}.

- periodic magnetic structures can produce a magnetic field which
decreases exponentially far from the structure and such a field
can be used as mirrors for cold atoms with a non vanishing
magnetic moment. For slow atoms, the angular momentum projection
$M$ is quantized on the magnetic field direction and follows
adiabatically the field, thus creating an attractive or repulsive
potential. This idea \cite{opat92,opat99} has been demonstrated by
several experiments \cite{roach95,drndic99}.

The coherence of the reflected wave depends on the mirror
roughness. Several experiments \cite{savalli2002,esteve04} have
tested the roughness of atomic mirrors: it is possible but
difficult to produce very coherent reflection.

Many atom traps (too numerous to be listed here) have been
developed for the production of quantum degenerate gases: these
traps rely either on magnetic forces or on the dipole potential in
far off-resonance laser beams. In an excellent vacuum (near
$10^{-10}$ mbar), the atom residence time in such a trap can be
quite long, (of the order $100$ s), limited either by collisions
with the residual gas or, for a dense trapped gas, by dimer
formation by $3$-body collisions. The atom coherence time, which
is more difficult to measure, is also sensitive to the
fluctuations of the trapping potential position and depth.

Miniature magnetic traps and waveguides are developed in order to
build integrated atom optics on a chip. In such traps where the
atom is very near a surface, new effects have been predicted by
Henkel \cite{henkel99a,henkel99b,henkel01}: the low-frequency part
of the thermal electromagnetic fields is considerably enhanced
near a conducting surface and these fields may reduce the
coherence lifetime in the trap. Some recent experiments have
observed the reduction of the atom residence time in magnetic
micro-traps when the atom-surface distance is reduced
\cite{fortagh02,jones03,rekdal04}. Finally, in 2005, after several
unsuccessful attempts, a Michelson atom interferometer has been
operated on a chip \cite{wang05}.

\section{Main types of atom interferometers}

\subsection{Polarization interferometers versus separated beam interferometers}

With light waves \cite{born75}, a distinction is classically made
between polarization interferometers (made of a polarizer, a
birefringent medium and an analyzer) and interferometers using
division of wavefront or of amplitude (in which a light beam is
split in two beams which recombine on the detector). Obviously,
this distinction is more technical than fundamental. Among atom
interferometers, all those using an inelastic diffraction process
(Ramsey-Bord\'e or Raman process) have a mixed character: the wave
follows two different paths but the beam-splitters have modified
the atom internal state.

With atoms, the equivalent of pure polarization interferometers
can be found in the Ramsey \cite{ramsey50} or Sokolov
\cite{sokolov73} experiments and in atomic clocks. The two paths
followed by the atom differ essentially by the internal states of
the atom and the momentum transfer, which is due to the absorption
of a microwave photon, is very small although not completely
vanishing \cite{wolf04}. Recent developments on polarization
interferometers, done by the group of Baudon and coworkers, are
reviewed in \cite{baudon99}. Here, we will concentrate on
interferometers in which the two atomic paths are noticeably
different.

\subsection{Interferometer designs}

With atomic waves, a complete equivalent of the Fabry-Perot or
Michelson interferometers is not feasible and, if we except some
Young's double slit type experiments, most interferometers are
based on the Mach-Zehnder design, with a diffraction process
replacing the mirrors and the beam-splitters. The high symmetry of
the Mach-Zehnder interferometer is very helpful to minimize the
sensitivity to defects but less symmetrical designs are also
interesting (see Fig. \ref{f3}). The Mach-Zehnder design can be
divided in various subtypes:

\begin{figure}
\includegraphics[width = 8 cm,height= 6cm]{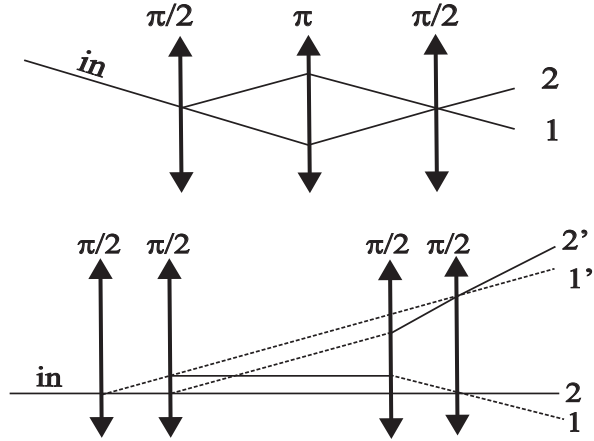}
\caption{\label{f3} Two examples of atom interferometers. The top
panel represents the simplest Mach-Zehnder interferometer, with
the input beam coming from the left and two complementary output
beams labelled $1$ and $2$ on the right. The lower panel shows an
asymmetrical atom interferometer with four diffraction processes
and four exit beams and such an interferometer can be used to
measure $h/M$ (see text). In both cases, the Rabi phases of the
diffraction processes are indicated, $\pi/2$ corresponding to an
ideal beam splitter and $\pi$ to a perfect mirror.}
\end{figure}

\begin{itemize}

\item temporal or spatial interferometers. In the simplest
temporal interferometers, the diffraction gratings are produced by
pulsing the laser beams at time $t=0$, $t=T$ and $t=2T$. In
spatial interferometers, three gratings, located at $z=0$, $z=L$
and $z=2L$, are successively crossed by the atoms. While spatial
interferometers are very similar to their traditional optics
counterparts, temporal interferometers, which are almost unknown
in optics, are easy to build with atoms thanks to laser
diffraction and, with cold atoms, temporal interferometers are the
usual choice.

In many cases, the phase shift to be measured varies with the time
interval $T$ and an accurate knowledge of $T$ is needed for a high
precision measurement. In a spatial interferometer, $T= L/v$ and
the dispersion of the phase shift due to the velocity distribution
limits the maximum observable phase shift and the accuracy of its
measurement. Several techniques \cite{hammond95,roberts04} have
been proposed to overcome this difficulty.

\item among temporal interferometers, some of them are based on an
echo-type technique, as discussed by Sleator and co-workers
\cite{cahn97}. With this technique, the atomic source transverse
velocity distribution may be larger than one photon recoil
velocity, an advantage shared with the Talbot-Lau interferometers.

\item the diffraction process itself can be used in the far-field
regime (the Fraunhofer regime where the beams associated to the
various diffraction orders do not overlap) or the near-field
regime. Near-field diffraction is used in Talbot-Lau
interferometers \cite{brezger03}, which is the only possible
design with heavy molecules, because of their very small de
Broglie wavelength: many such experiments have been done by Arndt,
Zeilinger and co-workers \cite{brezger02,hackermuller03} but a
Talbot-Lau interferometer has been developed with thermal atoms by
the group of Clauser \cite{clauser94}.

\item although most interferometers involve only two atomic paths,
a multiple path interferometer has been built and operated by
Weitz and co-workers \cite{weitz96}.

\end{itemize}

\section{Applications of atom interferometry}

Let us now review the main applications of atom interferometry,
illustrating each case by some experimental results.

\subsection{Young double slit interferometers and related experiments}

A Young's double slit experiment was realized by Carnal and Mlynek
in 1991 using a supersonic beam of metastable helium
\cite{carnal91}. A similar experiment was made in 1992 with
ultra-cold metastable neon by Shimizu and co-workers
\cite{shimizu92a}, who also observed the shift of the fringe
position due to an inhomogeneous electric field \cite{shimizu92b}.
Mlynek and co-workers have also used a charged wire to build a
kind of Fresnel biprism interferometer \cite{nowak98,nowak99}.

By modulating the laser power density of an evanescent wave
mirror, Dalibard and co-workers were able to induce a phase
modulation of an atomic wave \cite{steane95} and they used this
process to build a temporal Young double slit interferometer
\cite{szriftgiser96}.

\subsection{Effects of an electric field}

The electric polarizability $\alpha$ is interesting to measure by
atom interferometry, because spectroscopy can measure only the
polarizability difference between internal states. The phase shift
due to an electric field ${\mathbf E}$ applied on the
interferometer is given by:

\begin{equation}
\label{n3} \Delta \phi = \frac{2 \pi \epsilon_0 \alpha}{\hbar v}
\oint {\mathbf E}^2(s) ds
\end{equation}

\noindent where the path integral is taken on a closed circuit
following the two atomic paths inside the interferometer and $v$
is the atom velocity. It is necessary to apply an electric field
on only one of the interfering beams and this is possible, if the
collimation is sufficient, by using a capacitor with a thin
electrode (a septum) inserted between the two beams. With such an
experiment, Pritchard and co-workers \cite{ekstrom95} have
obtained a very accurate measurement of the sodium atom
polarizability. With a similar experiment, Toennies and co-workers
have compared the electric polarisabilities of helium atom and
dimer \cite{toennies03} and our group has measured the electric
polarizability of lithium atom \cite{miffre05a,miffre05b}. The
velocity dependence of the phase shift has been compensated by
applying time dependent phase shifts by Pritchard and co-workers
\cite{roberts04}.

With an interferometer using inelastic diffraction, a homogeneous
electric field applied on the two interfering beams induces a
phase shift proportional to the polarizability difference: such an
experiment was done on magnesium \cite{rieger93} and on calcium
\cite{morinaga96b}.

With a multiple beam interferometer, Weitz and co-workers have
observed the tensorial character of the phase shift due to the AC
Stark effect of a pulsed optical field \cite{weitz00}.

The Aharonov-Casher phase \cite{aharonov84}, which results from
the application of an electric field on an atom with an oriented
magnetic moment, has been measured by several experiments
\cite{sangster93,sangster95,zeiske95,yanagimachi02}.

\subsection{Effect of a magnetic field}

With paramagnetic atoms in a given $F, M_F$ hyperfine sublevel,
the phase shift due to the magnetic field $B$ is given by :
\begin{equation}
\label{n4} \Delta \phi(F,M_F)  = \frac{g_F \mu_B M_F}{\hbar v}
\oint B(s) ds
\end{equation}
\noindent where the field is assumed to vary slowly enough to
insure an adiabatic behaviour. The resulting phase shift $\Delta
\phi(F,M_F)$ vanishes for a homogeneous field and is proportional
to the magnetic field gradient. This gradient may be created by a
current sheet circulating between the two atomic beams
\cite{schmiedmayer94,schmiedmayer97} or more simply by a coil
\cite{giltner96,miffre05} or a wire \cite{wang05} (in this last
case, the gradient is pulsed).

In any case, it is difficult to evaluate the field integral very
accurately, so that these experiments cannot provide competitive
measurements of the Zeeman effect. In the experiments
\cite{schmiedmayer94,schmiedmayer97,giltner96,miffre05}, the
fringe visibility ${\mathcal{V}}$, which was recorded as a
function of the applied gradient, presents a series of revivals
when all the $\Delta \phi(F,M_F)$ are multiples of $2\pi$.

If the diffraction process is inelastic, a phase shift appears as
soon as the magnetic moments in the two states are not equal. If
the magnetic field is homogeneous and pulsed in time, the phase
shift is non-dispersive (i.e. independent of the atom velocity)
and this effect is a particular case of the scalar Aharonov-Bohm
effect. This effect has been studied on sodium by Morinaga and
co-workers \cite{shinohara02}. As a weak magnetic field induces a
large phase shift, most experiments try to cancel the sensitivity
to the magnetic field by pumping the atom in a $M_F= 0$ sub-level
(which has only a quadratic Zeeman effect) and by keeping the
magnetic field weak and homogeneous but non zero, as in atomic
clocks.

\subsection{Measurement of $h/M$ and of the fine structure constant $\alpha$}

With a proper design (see Fig. \ref{f3}), the interference signal
is sensitive to the photon recoil energy $\hbar \omega_{rec} =
\hbar^2 k_L^2/(2 M)$. The associated phase shift, which may be
very large, can be measured only with temporal interferometers
operated with cold atoms. The knowledge of $\hbar/M$, combined
with the Rydberg constant and mass ratios which are very
accurately known, gives access to the fine structure constant
$\alpha$. This new measurement is very interesting because it is
almost completely independent of quantum electrodynamics (QED)
calculations, while the best measurement of $\alpha$, deduced from
the anomalous Land\'e factor of the electron \cite{mohr05}, rely
on very complex QED calculations.

The first demonstration was made on cesium by Chu and co-workers
\cite{weiss93,weiss94} with an uncertainty of $0.1$ ppm on
$\hbar/M$ but their result was lower than the accepted value by
$0.85$ ppm. Many improvements have been made and, in 2002, this
experiment has given a measurement of $\alpha$ with an accuracy of
$7.4$ ppb \cite{wicht02}.

Similar experiments have been started on rubidium \cite{cahn97}
and hydrogen \cite{heupel02}. The contrast interferometer of
Pritchard and co-workers \cite{gupta02}, which operates with a
sodium BEC, has given a precision of $7$ ppm on $h/M_{Na}$, but
the result differs by $200$ ppm from the accepted value, a
difference attributed to the mean-field interaction in the
condensate. A recent experiment \cite{lecoq05} by Aspect and
co-workers with a rubidium BEC has observed a similar shift which
has been quantitatively related to the mean-field interaction in
the expanding BEC.

Bloch oscillations of an atom in a standing light wave were first
observed by the research groups of Salomon \cite{bendahan96} and
Raizen \cite{wilkinson96} in 1996. Such an experiment can give a
measurement of $h/M$ by measuring the transferred momentum through
a velocity measurement. A first experiment on rubidium $^{87}Rb$
has been made by Biraben and co-workers \cite{battesti04}, with an
uncertainty on $\alpha$ equal to $0.4$ ppm and this uncertainty
has been recently reduced to $6.7$ ppb \cite{clade06}.

\subsection{Measurement of inertial effects}

The sensitivity of a matter wave interferometer to inertial
effects is large \cite{anandan77,clauser88,borde89} as first
illustrated by the observation of the effect of gravity on a
neutron interferometer in 1975 \cite{colella75}. The classical
aspects of the detection of inertial effects with an atomic beam
have been discussed by the group of Zeilinger \cite{oberthaler96}.

\subsubsection{Measurement of accelerations}

The phase-shift due to an acceleration  ${\mathbf a}$ is given by:

\begin{equation}
\label{n5} \Delta\phi ={\mathbf k}_{eff} \cdot {\mathbf a} T^2
\end{equation}

\noindent where $T$ is the time between laser diffraction pulses
and $\hbar k_{eff}$ the momentum transferred to the atom by the
diffraction process. In the case of the acceleration of gravity
${\mathbf a} = {\mathbf g}$, minor corrections due to the gravity
gradient must be taken into account \cite{wolf99,peters01}.

The first measurement of ${\mathbf g}$ by atom interferometry was
performed in 1991 by Kasevich and Chu
\cite{kasevich91,kasevich92}, with a temporal interferometer with
Raman diffraction and laser cooled atoms. In 1997, Sleator and
co-workers \cite{cahn97} realized a preliminary measurement of $g$
with their echo interferometer. After a series of improvements,
Chu and co-workers \cite{peters99,peters01} have obtained an
uncertainty of $3\times 10^{-9}$ on $g$.

A gravity gradiometer has been built by the research group of
Kasevich \cite{snadden98,mcguirk02}, with an achieved sensitivity
equal to $4\times 10^{-9}$ s$^{-2}$ and this apparatus has been
used for a preliminary measurement of the gravitational constant
$G$ \cite{kasevich02}. Tino and co-workers are presently building
an atom interferometer dedicated to the measurement of $G$, aiming
at a $100$ ppm accuracy \cite{tino02,fattori03}.

An atom interferometer using amplitude gratings made of laser
standing waves has been built by H\"ansch and co-workers and it
has been used to test the equivalence principle with a $10^{-7}$
sensitivity \cite{fray04}.

The period of Bloch oscillations in a laser standing wave in the
vertical direction is directly related to $g$ and several
experiments have used this property to measure $g$. Inguscio and
co-workers \cite{roati04} have compared two realizations of this
experiment, one with a bosonic atom $^{87}$Rb and one with a
fermionic atom $^{40}$K, thus proving the superiority of
noninteracting fermions for such an experiment. Biraben and
co-workers \cite{clade05} have made a preliminary measurement of
$g$ with a $10^{-6}$ accuracy and pointed out several interesting
features of this technique.

If an atom interferometer is placed in an homogeneous electric
field, the atoms will be accelerated if their electric charge does
not exactly vanish. Chu and co-workers \cite{young97} proposed to
use the large sensitivity to accelerations of atom interferometers
to test the charge neutrality of atoms. Following
\cite{delhuille01}, it is possible to achieve a sensitivity near
$10^{-21}q_e$ ($q_e$: electron charge), equal to the sensitivity
achieved by experiments with neutrons \cite{baumann88} or
molecules \cite{dylla73}.

\subsubsection{Measurement of rotations: atom gyros}

The sensitivity of an interferometer to rotations is due to the
Sagnac effect. We can use equation (\ref{n4}) and replace the
acceleration ${\mathbf a}$  by the Coriolis term. Classically, the
phase-shift is  given by:

\begin{equation}
\label{n6} \Delta\phi = \frac{4 \pi {\mathbf \Omega} \cdot
{\mathbf A}}{\lambda_{dB} v}
\end{equation}

\noindent ${\mathbf \Omega}$ is the angular velocity of the
interferometer, ${\mathbf A}$ is the area enclosed by the
interferometer paths (normal to its surface) and $\lambda_{dB}$ is
the de Broglie wavelength . As this phase shift is an inertial
effect, it must be independent of the atom mass and this appears
obviously if one writes:

\begin{equation}
\label{n7}\Delta\phi = 2 \Omega v T^2 k_{eff} = 2 \Omega L^2
k_{eff}/v
\end{equation}

\noindent where $v$ is the atom velocity, $T$ the time between
diffraction pulses or $L$ the distance between diffraction
gratings. The exact value of the area $A$ is not a simple question
\cite{antoine03,antoine05}, because of the pulse duration or
spatial width of the laser beams.

The first atom interferometer gyrometer was built by Helmcke,
Bord\'e and co-workers \cite{riehle91} in 1991: it was a spatial
Ramsey-Bord\'e interferometer using a thermal calcium beam. A
spatial interferometer with a thermal sodium beam and material
gratings, built by Pritchard and co-workers \cite{lenef97}, has
reached a sensitivity equal to $3\times 10^{-6}$ rad/s$\sqrt{Hz}$.
A thermal cesium spatial interferometer with Raman diffraction by
Kasevich and co-workers \cite{gustavson97,gustavson00} has reached
a sensitivity equal to $6\times 10^{-10}$ rad/s$\sqrt{Hz}$. A
temporal interferometer using Raman diffraction with a cold cesium
fountain beam has been developed at BNM-SYRTE laboratory
\cite{leduc04} and a somewhat similar apparatus aiming at a very
high sensitivity needed for a direct detection of the
Lense-Thirring effect (a frame-dragging effect predicted by
general relativity \cite{ciufolini04}) is under development in a
joint effort involving several European laboratories (HYPER
project \cite{landragin01}).

\subsection{Other interactions}

Up to now, we have considered that the atoms inside the
interferometers are isolated and submitted only to electromagnetic
or inertial fields. In the following experiments, the atom
interactions are more complex, usually with a stochastic or
dissipative character.

\subsubsection{Effect of a gas on atomic wave propagation: index of refraction
or decoherence}

Pritchard and co-workers have measured the index of refraction of
various gases for sodium atomic waves
\cite{schmiedmayer95,schmiedmayer97,roberts02}. Using a gas cell
with a septum, a small gas density (corresponding to a gas
pressure near $10^{-3}$ millibar) is introduced on one of the
interfering beams (the atomic wave goes in and out of the cell
through thin slits). The phase shift and attenuation of the
transmitted wave are both detected on the interferometer signal.
In the case of rare gases, the experimental results can be
compared to calculations using the sodium-rare gas interaction
potentials and the main features are well explained
\cite{schmiedmayer95,champenois97,forrey96}.

The presence of a gas in the interferometer can also induce a
decoherence effect: this effect has been observed
\cite{hornberger03a,hornberger03b,hackermuller03a} in the
Talbot-Lau interferometer developed by Arndt, Zeilinger and
co-workers. The C$_{70}$ molecules are very massive and a
collision with an atom destroys the coherence of the spatial
wavefunction, by transferring some momentum to the C$_{70}$
molecules, but the resulting deviation is small enough so that the
molecules still arrive on the detector: as the coherence is lost,
the fringe visibility decrease with increasing gas density.

\subsubsection{Atom-surface van der Waals interaction}

When an atom goes through the narrow slits of a grating, the
atom-surface interaction is quite large and, even at thermal
velocities, this interaction cannot be neglected. Each slit can be
viewed as a cylindrical lens, with a velocity dependent focal
length. This effect modifies the diffraction amplitudes and it has
been studied carefully by the group of Toennies \cite{grisenti99}.
Moreover, as predicted by our group \cite{champenois99}, the
transmitted wave receives a phase shift which has been recently
measured by Cronin and co-workers \cite{perreault05}.

\subsubsection{Decoherence effects by spontaneous photon emission}

If an atom is excited by a laser, a spontaneous emission of a
photon occurs. This process transfers momentum to the atom and at
the same time gives a signature of the presence of the atom. Such
a process is therefore associated to a loss of coherence and the
visibility of the interference signals is reduced when spontaneous
photon emission occurs inside the interferometer.

This effect has been first studied by Mlynek and co-workers on the
laser diffraction pattern of a metastable helium beam
\cite{pfau94}. Then, Pritchard and co-workers were able to study
in great detail the loss of coherence due to the spontaneous
emission of a variable number of photons as a function of the
distance between the two interfering paths
\cite{chapman95,kokorowski01}. Mei and Weitz \cite{mei01,mei01a}
have extended this study, using their multiple path
interferometer, and they have shown that, in some cases,
decoherence can increase the fringe visibility.

Finally, an experiment due to Arndt, Zeilinger and co-workers
\cite{hackermuller04} investigates the same effect on C$_{70}$
molecules in their Talbot-Lau interferometer: the molecules,
heated by a laser before entering the interferometer, emit
infra-red photons and this emission induces decoherence.

\subsubsection{Decoherence effects by gravitational waves and related
topics}

Several works \cite{percival97,benatti02,chiao03}, some being
highly speculative, have discussed the detection of space-time
fluctuations with atom interferometers. The particular case of the
interaction of low-frequency gravitational waves with a
matter-wave interferometer can be described without any
approximation: a very intense background radiation of
gravitational waves emitted by binary star systems is predicted by
general relativity and its existence is commonly accepted. Because
there is a very large gain of sensitivity for inertial effects
when going from light wave to matter wave interferometers, the
same gain was expected for the detection of gravitational waves.
S. Reynaud and co-workers \cite{lamine02} have shown that this
gain of sensitivity does not exist.

\section{Conclusion}

This paper has given an overview of atom interferometry, limited
to interferometers in which the two atomic paths are spatially
different. We have also described the main applications of this
technique. We have quoted most of the pioneering works as well as
a large fraction of recent papers, illustrating the present state
of art, but many interesting papers have been omitted because of
lack of space. Let us summarize the main messages of this paper:

- atom interferometry has been rapidly expanding since 1991 and a
wide variety of experiments have already been realized. These
experiments give new access to the measurements of very different
quantities (atomic properties, accelerations and rotations,
fundamental constants $\alpha$ and $G$, quantum decoherence
effects, etc). Even if this list is already large, new
applications are still ahead!

- the possibilities opened by lasers to cool and manipulate
coherently atoms are extremely wide. In particular, the atom
internal states, which replace the polarization states of light,
give a considerably larger set of possibilities and this property
explains why so many different types of experiments can be
developed.

- atom interferometry has already achieved an extraordinary
sensitivity and many improvements are expected to provide further
gains of sensitivity. Better sources of atomic waves and, in
particular, the development of intense and continuous atom-lasers
will provide extraordinary improvements. The atom-atom
interactions, which give very impressive effects in quantum
degenerate gases, will then play an important role which has no
equivalent in traditional interferometry with photons.




\newpage
\begin{thebibliography}{}

\bibitem{carnal91} O. Carnal and J. Mlynek, Phys. Rev. Lett. {\bf 66},
 2689 (1991)

\bibitem{keith91} D. W. Keith, C. R. Ekstrom, Q. A. Turchette and D.
E. Pritchard, Phys. Rev. Lett. {\bf 66}, 2693 (1991)

\bibitem{riehle91} F. Riehle, Th. Kisters, A. Witte, J. Helmcke
and Ch. J. Bord\'e, Phys. Rev. Lett. {\bf 67}, 177 (1991)

\bibitem{kasevich91} M. Kasevich and S. Chu, Phys. Rev. Lett.
{\bf 67}, 181 (1991)

\bibitem{adams94} C. S. Adams, M. Sigel and J. Mlynek,
Phys. Rep. {\bf 240}, 143 (1994)

\bibitem{baudon99} J. Baudon, R. Mathevet and J. Robert,
J. Phys. B {\bf 32}, R173 (1999)

\bibitem{berman97} Atom interferometry edited by P. R. Berman
(Academic Press, 1997)

\bibitem{bongs04} K. Bongs and K. Sengstock, Rep. Prog. Phys. {\bf
67}, 907 (2004)

\bibitem{itano93} W. Itano J. C. Bergquist, J. J. Bollinger,
J. M. Gilligan, D. J. Heinzen, F. L. Moore, M. G. Raizen and D. J.
Wineland, Phys. Rev. A {\bf 47}, 3554 (1993)

\bibitem{santarelli99} G. Santarelli, Ph. Laurent, P. Lemonde, and
A. Clairon., Phys. Rev. Lett. {\bf 82}, 4619 (1999)

\bibitem{stern29} O. Stern, Naturwiss. {\bf 17}, 391 (1929)

\bibitem{estermann30} L. Estermann and O. Stern, Zeits. f. Physik,
{\bf 61}, 95 (1930)

\bibitem{ekstrom92} C. R. Ekstrom, D. W. Keith and D. E. Pritchard,
Appl. Phys. B {\bf 54}, 369 (1992)

\bibitem{savas95} T. A. Savas, S. N. Shah, M. L. Schattenburg, J. M.
Carter and H. I. Smith, J. Vac. Sci. Technol. B {\bf 13}, 2732
(1995)

\bibitem{schollkopf04} W. Sch\"ollkopf, R. E. Grisenti and J. P.
Toennies, Eur. Phys. J. D {\bf 28}, 125 (2004)

\bibitem{schollkopf94} W. Sch\"ollkopf and J. P. Toennies,
Science {\bf 266}, 1345 (1994)

\bibitem{morinaga96a} M. Morinaga, M. Yasuda, T. Kishimoto and F.
Shimizu , Phys. Rev. Lett. {\bf 77}, 802 (1996)

\bibitem{fujita00} J. Fujita, S. Mitake, F. Shimizu, Phys. Rev.
Lett. {\bf 84}, 4027 (2000)

\bibitem{shimizu02} F. Shimizu and J. I. Fujita, Phys. Rev. Lett. {\bf
88}, 123201 (2002).

\bibitem{kapitza33} P. L. Kapitza and P. A. M. Dirac, Proc.
Camb. Phil. Soc., {\bf 29}, 297-300 (1933)

\bibitem{batelaan01} D. L. Freimund, K. Aflatooni and H. Batelaan,
Nature {\bf 413} 142-143 (2001)

\bibitem{altshuler66} S. Altshuler, L. M. Frantz and S. Braunstein,
Phys. Rev. Lett., {\bf 17}, 231 (1966)

\bibitem{rolston02} S. L. Rolston and W. D. Philips, Nature
{\bf 416}, 219 (2002)

\bibitem{bernhardt81} A. F. Bernhardt and B. W. Shore, Phys. Rev.
{\bf A 23}, 1290 (1981)

\bibitem{giltner95a} D. M. Giltner, R. W. McGowan and Siu Au Lee,
Phys. Rev., {\bf A 52}, 3966 (1995)

\bibitem{keller99} C. Keller, J. Schmiedmayer, A. Zeilinger,
T. Nonn, S. D\"urr and G. Rempe, Appl. Phys. B {\bf 69}, 303
(1999)

\bibitem{letokhov81} V. S. Letokhov and V. G. Minogin, Phys.
Reports {\bf 73}, 1 (1981)

\bibitem{castin91} Y. Castin and J. Dalibard, Europhys. Lett. {\bf
14}, 761 (1991)

\bibitem{champenois01} C. Champenois, M. B\"uchner, R. Delhuille,
R. Mathevet, C. Robilliard, C. Rizzo and J. Vigu\'e, Eur. Phys. J.
D {\bf 13}, 271 (2001)

\bibitem{hall76} J. Hall, C. J. Bord\'e and K. Uehara, Phys. Rev.
Lett. {\bf 37}, 1339 (1976)

\bibitem{moskowitz83} P. E. Moskowitz, P. L. Gould, S. R. Atlas and
D. E. Pritchard, Phys. Rev. Lett. {\bf 51}, 370 (1983)

\bibitem{martin88} P. J. Martin, B. G. Oldaker, A. H. Miklich and D.
E. Pritchard, Phys. Rev. Lett., {\bf 60}, 515 (1988)

\bibitem{giltner95b} D. M. Giltner, R. W. McGowan and Siu Au Lee,
Phys. Rev. Lett. {\bf 75}, 2638 (1995)

\bibitem{borde89} C. J. Bord\'e, Phys. Lett. A {\bf 140}, 10 (1989)

\bibitem{borde97} C. J. Bord\'e, Atom interferometry, edited by P.
R. Berman, (Academic Press, 1997) p. 257

\bibitem{marte91} P. Marte, P. Zoller and J. L. Hall, Phys. Rev. A
{\bf 44}, R4118 (1991)

\bibitem{weitz94} M. Weitz, B. C. Young and S. Chu, Phys. Rev.
Lett. {\bf 73}, 2563 (1994) and Phys. Rev. A {\bf 50}, 2438 (1994)

\bibitem{aminoff93} C. G. Aminoff, A. M. Steane, P. Bouyer, P. Desbiolles,
J. Dalibard and C. Cohen-Tannoudji, Phys. Rev. Lett. {\bf 71},
3083 (1993)

\bibitem{landragin96}A. Landragin, J.-Y. Courtois, G. Labeyrie,
N. Vansteenkiste, C. I. Westbrook and A. Aspect, Phys. Rev. Lett.
{\bf 77}, 1464 (1996)

\bibitem{opat92} G. I. Opat, S. J. Wark and A. Cimmino, Appl.
Phys. B {\bf 54}, 396 (1992)

\bibitem{opat99} G. I. Opat, S. Nis Chormaic, B. P. cantwell and J. A.
Richmond, J. Opt. B: Quantum Semiclass. Opt. {\bf 1}, 415 (1999)

\bibitem{roach95} T. M. Roach, H. Abele, M. G. Boshier, H. L. Grossman,
K. P. Zetie and E. A. Hinds, Phys. Rev. Lett. {\bf 75}, 629 (1995)

\bibitem{drndic99} M. Drndic, G. Zabow, C. S. Lee, J. H. Thywissen,
K. S. Johnson, M. Prentiss, and R. M. Westerveltet al., Phys. Rev.
A {\bf 60}, 4012 (1999).

\bibitem{savalli2002} V. Savalli, D. Stevens, J. Estève, P. D. Featonby,
V. Josse, N. Westbrook, C. I. Westbrook and A. Aspect, Phys. Rev.
Lett. {\bf 88}, 250404 (2002)

\bibitem{esteve04} J. Est\`eve, D. Stevens, C. Aussibal, N. Westbrook,
A. Aspect and C. I. Westbrook, Eur. Phys. J. D {\bf 31}, 487
(2004)

\bibitem{henkel99a} C. Henkel and M. Wilkens, Europhys. Lett. {\bf
47}, 414 (1999)

\bibitem{henkel99b} C. Henkel, S. P\"otting and M. Wilkens, Appl.
Phys. B {\bf 69}, 379 (1999)

\bibitem{henkel01} C. Henkel and S. P\"otting, Appl. Phys. B
{\bf 72}, 73 (2001)

\bibitem{fortagh02} J. Fortagh, H. Ott, S. Kraft, A. Günther
and C. Zimmermann, Phys. Rev. A {\bf 66}, 41604(R) (2002)

\bibitem{jones03} M. P. A. Jones, C. J. Vale, D. Sahagun, B. V. Hall and E. A.
Hinds, Phys. Rev. Lett. {\bf 91}, 080401 (2003)

\bibitem{rekdal04} P. K. Rekdal, S. Scheel, P. L. Knight and E. A. Hinds,
Phys. Rev. A {\bf 70}, 013811 (2004)

\bibitem{wang05} Y.-J. Wang, D. Z. Anderson, V. M. Bright, E. A. Cornell,
Q. Diot, T. Kishimoto, M. Prentiss, R. A. Saravanan, S. R. Segal
and S. Wu, Phys. Rev. Lett. {\bf 94}, 090405 (2005)

\bibitem{born75} M. Born and E. Wolf, Principle of Optics,
Pergamon Press (1975)

\bibitem{ramsey50} N. F. Ramsey, Phys. Rev. {\bf 78}, 795 (1950)

\bibitem{sokolov73} Y. L. Sokolov, Sov. Phys. JETP, {\bf 36}, 243
(1973)

\bibitem{wolf04} P. Wolf and C. J. Bord\'e, arXiv
quant-ph/0403194

\bibitem{hammond95} T. D. Hammond, D. E. Pritchard, M. S. Chapman
A. Lenef and J. Schmiedmayer, Appl. Phys. B {\bf 60}, 193 (1995)

\bibitem{roberts04} T. D. Roberts, A. D. Cronin, M. V. Tiberg and
D. E. Pritchard, Phys. Rev. Lett. {\bf 92}, 060405 (2004)

\bibitem{cahn97} S. B. Cahn, A. Kumarakrishnan, U. Shim, and T. Sleator,
Phys. Rev. Lett. {\bf 79}, 784 (1997)

\bibitem{brezger03}B. Brezger, M. Arndt and A. Zeilinger, J. Opt. B:
Quantum Semiclass. Opt. {\bf 5}, S82 (2003)

\bibitem{brezger02} B. Brezger, L. Hackerm\"uller, S. Uttenthaler,
J. Petschinka, M. Arndt and A. Zeilinger, Phys. Rev. Lett. {\bf
88}, 100404 (2002)

\bibitem{hackermuller03} L. Hackerm\"uller, S. Uttenthaler, K. Hornberger,
E. Reiger, B. Brezger, A. Zeilinger and M. Arndt, Phys. Rev. Lett.
{\bf 91}, 090408 (2003)

\bibitem{clauser94} J. F. Clauser and Shifang Li, Phys. Rev. A
{\bf 49}, R2213 (1994) and Phys. Rev. A {\bf 50}, 2430 (1994)

\bibitem{weitz96} M. Weitz, T. Heupel and T. W. H\"ansch, Phys.
Rev. Lett. {\bf 77}, 2356 (1996)

\bibitem{shimizu92a} F. Shimizu, H. Shimizu and H. Takuma, Phys.
Rev. A {\bf 46}, R17 (1992)

\bibitem{shimizu92b} F. Shimizu, H. Shimizu and H. Takuma, Jpn. J.
Appl. Phys., {\bf 31}, L436 (1992)

\bibitem{nowak98} S. Nowak, N. Stuhler, T. Pfau and J. Mlynek, Phys. Rev. Lett.
{\bf 81}, 5792 (1998)

\bibitem{nowak99} S. Nowak, N. Stuhler, T. Pfau and J. Mlynek, Appl. Phys. B {\bf
69}, 269 (1999)

\bibitem{steane95} A. Steane, P. Szriftgiser, P. Desbiolles and J. Dalibard,
Phys. Rev. Lett. {\bf 74}, 4972 (1995)

\bibitem{szriftgiser96} P. Szriftgiser, D. Gu\'ery-Odelin, M. Arndt and J. Dalibard
Phys. Rev. Lett. {\bf 77}, 4 (1996)

\bibitem{ekstrom95} C. R. Ekstrom, J. Schmiedmayer, M. S. Chapman,
T. D. Hammond and D. E. Pritchard, Phys. Rev. A {\bf 51}, 3883
(1995)

\bibitem{toennies03} J. P. Toennies, private communication (2003)

\bibitem{miffre05a} A. Miffre, M. Jacquey, M. B\"uchner, G. Tr\'enec and J. Vigu\'e,
 Phys. Rev. A 73, 011603(R) (2006)

\bibitem{miffre05b} A. Miffre, M. Jacquey, M. B\"uchner, G. Tr\'enec and J.
Vigu\'e, Eur. Phys. J. D. {\bf 38}, 353 (2006)

\bibitem{rieger93} V. Rieger, K. Sengstock, U. Sterr, J. H. M\"uller
and W. Ertmer, Optics Com. {\bf 99}, 172 (1993)

\bibitem{morinaga96b} A. Morinaga, M. Nakalmura, T. Kuorosu and N.
Ito, Phys. Rev. A {bf 54}, R21 (1996)

\bibitem{weitz00} M. Mei, T. W. H\"ansch and M. Weitz, Phys. Rev.
A {\bf 61}, 020101(R) (2000)

\bibitem{aharonov84} Y. Aharonov and A. Casher, Phys. Rev. Lett.
{\bf 53}, 319 (1984)

\bibitem{sangster93} K. Sangster, E. A. Hinds, S. M. Barnett and E.
Riis, Phys. Rev. Lett. {\bf 71}, 3641 (1993)

\bibitem{sangster95} K. Sangster, E. A. Hinds, S. M. Barnett, E.
Riis and A. G. Sinclair, Phys. Rev. A {\bf 51}, 1776 (1995)

\bibitem{zeiske95} K. Zeiske, G. Zinner, F. Riehle and J. Helmcke,
Appl. Phys. B {\bf 60}, 295 (1995)

\bibitem{yanagimachi02} S. Yanagimachi, M. Kajiro, M. Machiya
and A. Morinaga, Phys. Rev. A {\bf65}, 042104 (2002)

\bibitem{schmiedmayer94} J. Schmiedmayer, C. R. Ekstrom, M. S.
Chapman, T. D. Hammond and D. E. Pritchard, J. Phys. II France
{\bf 4}, 2029 (1994)

\bibitem{schmiedmayer97} J. Schmiedmayer, M. S. Chapman, C. R. Ekstrom,
T. D. Hammond, D. A. Kokorowski, A. Lenef, R. A Rubenstein, E. T.
Smith and D. E. Pritchard in Atom interferometry edited by P. R.
Berman (Academic Press, 1997), p. 1

\bibitem{giltner96} D. M. Giltner, Ph. D. thesis, Colorado State
University, Fort Collins (1996)

\bibitem{miffre05} A. Miffre, M. Jacquey, M. B\"uchner, G. Tr\'enec and J.
Vigu\'e, Eur. Phys. J. D {\bf 33}, 99 (2005)

\bibitem{shinohara02} K. Shinohara, T. Aoki and A. Morinaga, Phys.
Rev. A {\bf 66}, 042106 (2002)

\bibitem{mohr05} P. J. Mohr and B. N. Taylor, Rev. Mod. Phys. {\bf
77}, 1 (2005)

\bibitem{weiss93} D. S. Weiss, B. C. Young and S. Chu, Phys. Rev. Lett.
{\bf 73}, 2706 (1993)

\bibitem{weiss94} D. S. Weiss, B. C. Young and S. Chu, Appl. Phys.
B {\bf 59}, 217 (1994)

\bibitem{wicht02} A. Wicht, J. M. Hensley, E. Sarajlic and S. Chu,
Physica Scripta, {\bf T1002}, 82 (2002)

\bibitem{heupel02} T. Heupel, M. Mei, M. Niering, B. Gross,
M. Weitz, T. W. H\"ansch and C. J. Bord\'e, Europhys. Lett. {\bf
57}, 158 (2002)

\bibitem{gupta02} S. Gupta, K. Dieckmann, Z. Hadzibabic and D. E.
Pritchard, Phys. Rev. Lett. {\bf 89}, 140401 (2002)

\bibitem{lecoq05} Y. Lecoq, J. A. Retter, S. Richard, A. Aspect
and P. Bouyer, submitted to Appl. Phys. B and
arXiv:cond-mat/0501520

\bibitem{bendahan96} M. Ben Dahan, E. Peik, J. Reichel, Y. Castin
and C. Salomon, Phys. Rev. Lett. {\bf 76}, 4508 (1996)

\bibitem{wilkinson96} S. R. Wilkinson, C. F. Bharucha, K. W. Madison,
Qian Niu and M. G. Raizen , Phys. Rev. Lett. {\bf 76}, 4512 (1996)

\bibitem{battesti04} R. Battesti, P. Clad\'e, S. Guellati-Kh\'elifa, C. Schwob,
B. Gr\'emaud, F. Nez, L. Julien and F. Biraben, Phys. Rev. Lett.
{\bf 92}, 253001 (2004)

\bibitem{clade06} P. Clad\'e, E. de Mirandes, M. Cadoret, S.
Guellati-Kh\'elifa, C. Schwob, F. Nez, L. Julien and F. Biraben,
Phys. Rev. Lett. {\bf 93}, 033001 (2006)

\bibitem{anandan77} J. Anandan, Phys. Rev. D {\bf 15}, 1448 (1977)

\bibitem{clauser88} J. F. Clauser, Physica B {\bf 151}, 262 (1988)

\bibitem{colella75} R. Colella, A. W. Overhauser, S. A. Werner,
Phys. Rev. Lett. {\bf 34}, 1472 (1975)

\bibitem{oberthaler96} M. K. Oberthaler, S. Bernet, E. M. Rasel, J.
Schmiedmayer and A. Zeilinger, Phys. Rev. A {\bf 54} 3165 (1996)

\bibitem{wolf99} P. Wolf anf P. Tourrenc, Phys. Lett. A {\bf 251},
241 (1999)

\bibitem{peters01} A. Peters, K. Y. Chung and S. Chu, Metrologia
{\bf 38}, 25 (2001)

\bibitem{kasevich92} M. Kasevich and S. Chu, Appl. Phys. B {\bf
54}, 321 (1992)

\bibitem{peters99} A. Peters, K. Y. Chung and S. Chu, Nature {\bf 400},
849 (1999)

\bibitem{snadden98} M. J. Snadden, J. M. McGuirk, P. Bouyer, K. G. Haritos
and M. A. Kasevich, Phys. Rev. Lett. {\bf 81}, 971 (1998)

\bibitem{mcguirk02} J. M. McGuirk, G. T. Foster, J. B. Fixler, M. J. Snadden
and M. A. Kasevich, Phys. Rev. A {\bf 65}, 033608 (2002)

\bibitem{kasevich02} M. A. Kasevich, communication to HYPER
symposium (Paris, November 2002)

\bibitem{tino02} G. M. Tino, Nucl. Phys. B {\bf 113}, 289 (2002)

\bibitem{fattori03} M. Fattori, G. Lamporesi, T. Petelski, J. Stuhler
and G. M. Tino, Phys. Lett. A {\bf 318}, 184 (2003)

\bibitem{fray04} S. Fray, C. Alvarez Diez, T. W. H\"ansch and M. Weitz ,
Phys. Rev. Lett. {\bf 93}, 240404 (2004)

\bibitem{roati04} G. Roati, E. de Mirandes, F. Ferlaino, H. Ott,
G. Modugno, and M. Inguscio, Phys. Rev. Lett. {\bf 92}, 230402
(2004)

\bibitem{clade05} P. Clad\'e, S. Guellati-Kh\'elifa, C. Schwob,
F. Nez, L. Julien and F. Biraben, Europhys. Lett. {\bf71}, 730
(2005)

\bibitem{young97} B. Young, M. Kasevich and S. Chu, in Atom
interferometry edited by P. R. Berman (Academic Press, 1997), p.
363

\bibitem{delhuille01} R. Delhuille, C. Champenois, M. B\"uchner, R.
Mathevet, C. Rizzo, C. Robilliard and J. Vigu\'e, QED 2000, G.
Cantatore ed., AIP Conference Proceedings {\bf 564}, 192 (2001)

\bibitem{baumann88} J. Baumann, R. G\"ahler, J. Kalus and W. Mampe,
Phys. Rev D {\bf 37}, 3107 (1988)

\bibitem{dylla73} H. F. Dylla and J. G. King, Phys. Rev. A {\bf
7}, 1224 (1973)

\bibitem{antoine03} C. Antoine and C. J. Bord\'e, Phys. Lett. A
{\bf 306}, 277 (2003)

\bibitem{antoine05} C. Antoine, Ph. D. thesis, Universit\'e P. et
M. Curie (2004)

\bibitem{lenef97} A. Lenef, T. D. Hammond, E. T. Smith, M. S. Chapman, R. A.
Rubenstein and D. E. Pritchard, Phys. Rev. Lett. {\bf 78}, 760
(1997)

\bibitem{gustavson97} T. L. Gustavson, P. Bouyer and M. A.
Kasevich, Phys. Rev. Lett. {\bf 78}, 2046 (1997)

\bibitem{gustavson00} T. L. Gustavson, A. Landragin and M. A.
Kasevich, Class. Quantum Grav. {\bf 17}, 2385 (2000)

\bibitem{leduc04} F. Leduc, D. Holleville, J. Fils, A. Clairon, N. Dimarcq
A. Landragin, P. Bouyer and C. J. Bord\'e, Proceedings of 16th
ICOLS, P. Hannaford et al. editors (World Scientific, 2004) p. 68

\bibitem{ciufolini04} I. Ciufolini and E. C. Pavlis, Nature {\bf 431},
958 (2004)

\bibitem{landragin01} A. Landragin, A. Clairon, N. Dimarcq,
P. Teyssandier, C. Salomon, E. M. Rasel, W. Ertmer, C. J. Bord\'e,
P. Tourrenc, P. Bouyer, M. Caldwell, R. Bingham, B. Kent, M.
Sandford, P. Wolf, S. Airey and G. Bagnasco, Proceedings of
ICATPP-7, M. Barone et al. editors (World Scientific, 2002) p. 16

\bibitem{schmiedmayer95} J. Schmiedmayer, M. S. Chapman, C. R. Ekstrom,
T. D. Hammond, S. Wehinger and D. E. Pritchard, Phys. Rev. Lett.
{\bf 74}, 1043 (1995)

\bibitem{roberts02} T. D. Roberts, A. D. Cronin, D. A. Kokorowski
and D. E. Pritchard, Phys. Rev. Lett. {\bf 89}, 200406 (2002)

\bibitem{champenois97} C. Champenois, E. Audouard, P. Duplaa and
J. Vigu\'e, J. Phys. II France {\bf 7}, 523 (1997)

\bibitem{forrey96} R. C. Forrey, Li You, V. Kharchenko and
A. Dalgarno, Phys. Rev. A {\bf 54}, 2180 (1996)

\bibitem{hornberger03a} K. Hornberger, S. Uttenthaler, B. Brezger,
L. Hackermüller, M. Arndt and A. Zeilinger , Phys. Rev. Lett. {\bf
90}, 160401 (2003)

\bibitem{hornberger03b} K. Hornberger and J. E. Sipe, Phys. Rev.
A {\bf 68}, 012105 (2003)

\bibitem{hackermuller03a} L. Hackermuller, K. Hornberger, B. Brezger,
A. Zeilinger and M. Arndt, Appl. Phys. B {\bf 77}, 781 (2003)

\bibitem{grisenti99} R. E. Grisenti, W. Sch\"ollkopf and J. P. Toennies,
Phys. Rev. Lett. {\bf 83}, 1755 (1999)

\bibitem{champenois99} C. Champenois, M. B\"uchner and J. Vigu\'e,
Eur. Phys. J. D {\bf 5}, 363 (1999).

\bibitem{perreault05} J. D. Perreault and A. D. Cronin, Phys. Rev.
Lett. {\bf 95}, 133201 (2005)

\bibitem{pfau94} T. Pfau, S. Sp\"alter, Ch. Kurtsiefer, C. R. Ekstrom
and J. Mlynek, Phys. Rev. Lett. {\bf 73}, 1223 (1994)

\bibitem{chapman95} M. S. Chapman, T. D. Hammond, A. Lenef, J. Schmiedmayer,
R. A. Rubenstein, E. Smith and D. E. Pritchard, Phys. Rev. Lett.
{\bf 75}, 3783 (1995)

\bibitem{kokorowski01} D. A. Kokorowski, A. D. Cronin, T. D. Roberts and
D. E. Pritchard, Phys. Rev. Lett. {\bf 86}, 2191 (2001)

\bibitem{mei01} M. Mei and M. Weitz, Phys. Rev. Lett. {\bf 86}, 559
(2001)

\bibitem{mei01a}  M. Mei and M. Weitz, Appl. Phys. B {\bf 72}, 91
(2001)

\bibitem{hackermuller04} L. Hackermuller, K. Hornberger, B. Brezger,
A. Zeilinger and M. Arndt, Nature {\bf 427}, 711 (2004)

\bibitem{percival97} I. C. Percival and W. T. Strunz, Proc.
Roy. Soc. A {\bf 453}, 431 (1997)

\bibitem{chiao03} R. Y. Chiao and A. D. Speliotopoulos,
Int. J. Mod. Phys. D {\bf 12}, 1627 (2003)

\bibitem{benatti02} F. Benatti and R. Floreanini, Phys. Rev. A
{\bf 66}, 043617 (2002)

\bibitem{lamine02} B. Lamine, M.-T. Jaekel and S. Reynaud, Eur. Phys. J. D {\bf 20}, 165
(2002)



\end{thebibliography}
\end{document}